\documentclass[letterpaper, 10 pt, conference]{ieeeconf}  

\IEEEoverridecommandlockouts                              

\makeatletter
\def\endthebibliography{%
	\def\@noitemerr{\@latex@warning{Empty `thebibliography' environment}}%
	\endlist
}
\makeatother

\usepackage{amsmath} 
\usepackage{amssymb}  
\usepackage[squaren, Gray, cdot]{SIunits}
\usepackage{cite}
\usepackage{graphicx}
\usepackage{subcaption} 

\title{\LARGE \bf
MagHT: a Magnetic Hough Transform for Fast Indoor Place Recognition
}

\author{Iad ABDUL RAOUF$^{1,2}$, Vincent GAY-BELLILE$^{1}$, Steve BOURGEOIS$^{1}$, Cyril JOLY$^{2}$ and Alexis PALJIC$^{2}$
\thanks{$^{1}$Authors are with Université Paris-Saclay, CEA, List, F-91120, Palaiseau, France {\tt\small iad.abdulraouf@cea.fr}}%
\thanks{$^{2}$Authors are with Centre de Robotique, MINES ParisTech, Université PSL, 75006 Paris, France {\tt\small iad.abdul\_raouf@mines-paristech.fr}}%
}

\begin{document}

\maketitle
\thispagestyle{empty}
\pagestyle{empty}

\begin{abstract}

This article proposes a novel indoor magnetic field-based place recognition algorithm that is accurate and fast to compute.
For that, we modified the generalized "Hough Transform" to process magnetic data (MagHT).
It takes as input a sequence of magnetic measures whose relative positions are recovered by an odometry system and recognizes the places in the magnetic map where they were acquired.
It also returns the global transformation from the coordinate frame of the input magnetic data to the magnetic map reference frame.
Experimental results on several real datasets in large indoor environments demonstrate that the obtained localization error, recall, and precision are similar to or are better than state-of-the-art methods while improving the runtime by several orders of magnitude.
Moreover, unlike magnetic sequence matching-based solutions such as DTW, our approach is independent of the path taken during the magnetic map creation.

\end{abstract}

\section{INTRODUCTION}

Place recognition is a popular solution for indoor localization due to the absence of GNSS signals. Vision-based approaches \cite{campos_orb-slam3_2021} are the most common. However, they fail in low or repetitive textured environments and when visual cues change as time goes by \cite{coulin_tightly-coupled_2022}. Some work improves the robustness, for instance, by learning different recognition modules for various illumination conditions \cite{labbe_multi-session_2022}. However, it tackles only one of the problems and increases the computation time.

On the contrary, many works have shown that the ambient magnetic field spatial variability can be used reliably for indoor localization over long periods \cite{coulin_tightly-coupled_2022,ouyang_survey_2022}. Indoor, the magnetic field varies slowly because of the natural variation of the global earth's magnetic field over the years \cite{ouyang_survey_2022}. It is also tolerant of moving objects in the scene. Human crowd or non-ferromagnetic furniture does not significantly impact the magnetic field \cite{ashraf_comprehensive_2020}. In contrast, larger objects such as a car and an elevator cabin modify it for a few meters \cite{shu_magicol_2015}.

Magnetic field-based relocalization algorithms exploit several magnetic data collected along a trajectory (called "input trajectory" thereafter), a single measure not being sufficiently discriminating to differentiate between places. Most existing approaches assume that those magnetic data are acquired by following the path taken to build the magnetic map \cite{antsfeld_magnetic_2021,lee_amid_2018,wang_keyframe_2016,yang_robust_2022}. Other probabilistic approaches \cite{solin_terrain_2016,ma_basmag_2016} do not constraint the movement but require heavy computational resources to achieve good performances (recognition rate, false positive rate, etc.) or reduced runtime is achieved through performance sacrifices.

This paper describes a new magnetic place recognition algorithm for large indoor environments. Unlike previous work, it is both fast to compute and makes no assumptions about the path taken. Consequently, it is suitable for 3D navigation in any environment, including large open areas. It only assumes prior reconstruction of the input trajectory in a gravity frame by an odometry system that includes an inertial measurement unit (IMU). The proposed approach adapts the generalized Hough transform \cite{ballard_generalizing_1981} to magnetic data. It results in several contributions, which are the following:
\begin{enumerate}
	\item A fast-matching method between input magnetic data and a magnetic map, robust to small spatial variations in the magnetic field.
	\item A voting process that estimates the transformation candidates between the global reference frame of the map and the input trajectory frame. The voting and the matching treat input magnetic data independently to ensure our method is independent of the mapping path.
	\item A pose estimation method to process all individual votes efficiently, robust to more than 99 \% outliers, that yields a unique estimation of the frame transformation.
	\item Overall, a place recognition algorithm that combines the previous contributions into a method several orders of magnitude faster than the state of the art while having similar or better recall, precision, and errors.
\end{enumerate}

To support our claims, we present experimental results of the proposed method inside open and closed multi-story indoor environments and an extensive comparison against a state-of-the-art method \cite{solin_terrain_2016}.

\section{RELATED WORK}

Magnetic place recognition algorithms identify where a collection of magnetic measures have been acquired in a magnetic map of the environment.

The first family of approaches relies exclusively on a temporal series of magnetic measures provided by a magnetometer. To reduce the problem's complexity, they assume that the set of possible trajectories is countable. Consequently, the magnetic map is reduced to a collection of temporal series of magnetic measures, and the place recognition to a pattern matching problem between temporal series. While the first solutions relied on handmade algorithms such as DTW \cite{wang_keyframe_2016,yang_robust_2022}, their too-high false positive rate requires their use to be combined with additional inputs, such as WIFI \cite{yang_robust_2022} or motion patterns \cite{yang_robust_2022}, to confirm the place recognition. This issue was solved more recently by deep-learning-based pattern matching \cite{antsfeld_magnetic_2021,lee_amid_2018}. However, all those solutions remain limited by their map representation. While the hypothesis of a limited number of trajectories is acceptable for corridor environments, such a hypothesis is no longer acceptable for open areas such as an atrium.

To tackle this limitation, the second family of approaches proposes to replace temporal pattern matching with spatial pattern matching. For that, an additional odometry system provides a spatial trajectory in addition to the magnetic measurements. The pattern matching is then achieved by exploring a spatial map of the magnetic field through particle filtering \cite{solin_terrain_2016} or Hidden Markov Model \cite{ma_basmag_2016}, the measurements being compared to the map prediction for the hypothesized locations. While those approaches are not limited to predetermined paths, they suffer from a computational time burden that increases with the size of the map.

To our knowledge, no algorithm can perform magnetic place recognition in large-scale environments independently from the mapping path and with a low computational cost. We designed a novel method described in section \ref{sec:maght} that satisfies all these conditions.

\section{MAGNETIC HOUGH TRANSFORM}\label{sec:maght}

\begin{figure*}
    \centering
    \includegraphics[trim=0 0 0 -10,clip,width=0.98\linewidth]{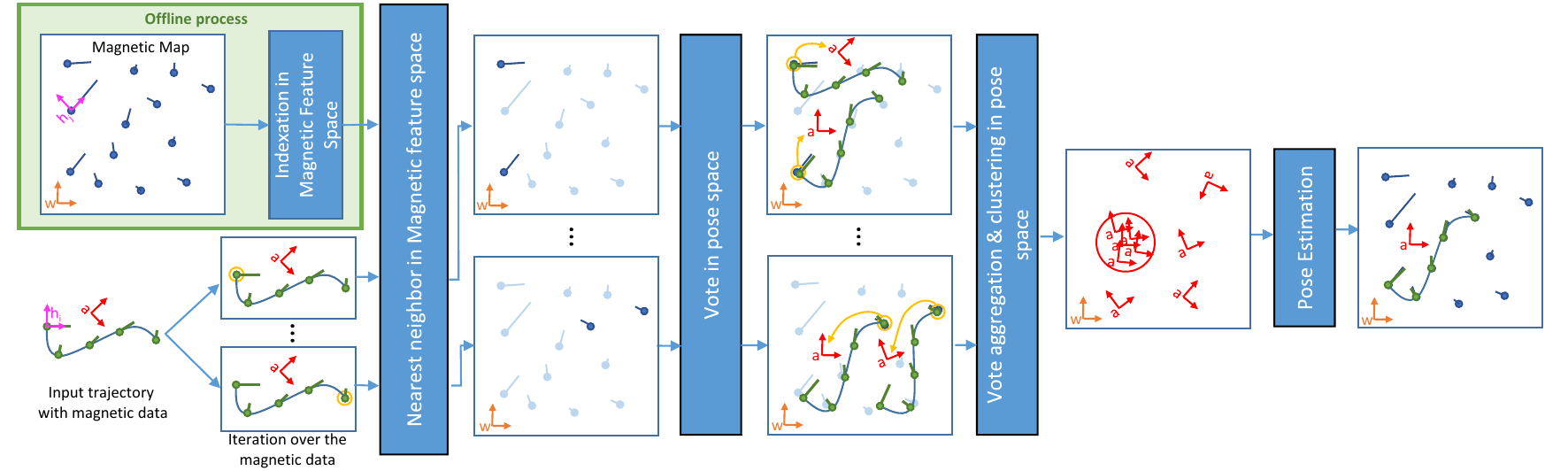}
    \caption{Inputs are a discrete magnetic map in a gravity frame $w$ (orange) and a trajectory with magnetic data expressed in another gravity frame $a$ (red). Each blue and green vector represents a magnetic vector of the map and a magnetic vector of the input trajectory. Frames $h_i$ and $h_j$ are gravity frames defined for each magnetic vector $m_i$ and $m_j$. Their $(Ox)$ axis are aligned with $m_{xy}$. Each measured magnetic vector $m_i$ matches multiples $m_j$ through the feature space. For each magnetic match, one vote is generated by the superimposition of their respective $h$ frame. After all input magnetic data have been handled, clustering search accumulations among all votes. The biggest cluster is selected to estimate the transformation $T_{wa}$ from its centroid.}
    \label{fig:overview}
\end{figure*}

\subsection{Overview}

Our goal is to find the rigid 3D transformation $T_{wa}\in SE(3)$ from the reference frame $a$ associated with a short input trajectory recovered by an odometry system (e.g., visual-inertial SLAM \cite{campos_orb-slam3_2021}) to a world reference frame $w$ associated to a magnetic map. A rigid transformation model supposes that the odometry drift is neglected, a reasonable assumption for short trajectories of a few meters. Frames a and w are also assumed to be gravity frames, meaning their z-axis is upright. This vertical alignment is obtained from an IMU sensing the gravity vector. It constrains $T_{wa}$ to have only 4 degrees of freedom $(x, y, z, \psi) \in \mathbb{R}^4$ with $t = (x, y, z)^\top$ the translation and $\psi$ the yaw angle. The estimation of $T_{wa}$ is computed by linking a finite set of magnetic vectors $m_j^w$ from the map with another finite set of measured magnetic vectors $m_{i}^a$ along the input trajectory.

For that, we used a generalized Hough transform. It is a method for converting data from an input space to a parameter space. Intuitively, instead of detecting the contour of an object in a 2D image by a visual Hough algorithm \cite{ballard_generalizing_1981}, our Magnetic Hough Transform (MagHT) detects any trajectory shape in a 3D magnetic map. It is composed of the following steps:

\begin{itemize}
	\item \textbf{Voting}: For each measures $m_i^a$, all compatible magnetic vectors $m_j^w$ are extracted to form several pairs. Then each pair leads to voting for one possible estimation $\tilde{T}_{wa}$ of the rigid transformation.
	\item \textbf{Clustering}: Depending on the size of the map, a few dozen input measurements can generate thousands of votes. The clustering step allows to identify a consensus around similar values.
	\item \textbf{Estimation}: The final $\hat{T}_{wa}$ estimate is taken equal to the centroid of the largest $\tilde{T}_{wa}$ cluster.
\end{itemize}

Each step is detailed in the following sections and illustrated in Fig \ref{fig:overview}.

\subsection{Pose Voting}\label{sec:vote}

The Hough transform first matches magnetic vectors from the input trajectory with similar magnetic vectors of the map to compute a vote $\tilde{T}_{wa}$.

\subsubsection{Feature matching} Because ${T}_{wa}$ is unknown, it is impossible to use the full magnetic vector whose expression depends on the orientation of its frame. However, both frames are gravity frames. Consequently, we use the following yaw invariant features $(\lVert m_{xy} \lVert, m_z) \in \mathbb{R_+} \times \mathbb{R}$, with $\lVert m_{xy} \lVert$ the norm of the horizontal component and $m_z$ the vertical component. Then each input's magnetic feature is associated with all map's magnetic features close enough (i.e., within a radius $\delta$) in the feature space $\mathbb{R_+} \times \mathbb{R}$. At places where the field is relatively uniform, $\delta$ should be small not to match too many features and vice versa, so we made it adaptative based on local input magnetic field variations (details are given in section \ref{sec:maghtSetup}). Computing all possible couple's distances to find the ones close enough is too expensive. Thereby map features are indexed into a k-d tree, well known to enable fast range search in small dimensions. 

\subsubsection{Voting}  Then, for each matched pair $(m_i^a, m_j^w)$, an estimation $\tilde{T}_{wa}$ is defined such that both magnetic vectors and their respective position are "aligned" while respecting the gravity constraint. Formally, this is done through magnetic frames $h_i$ and $h_j$ introduced in Fig. \ref{fig:overview}. Both are identified as the same frame $h$ created from the same physical reality (independent of $a$ or $w$ in which $h$ is defined). Thus a vote $\tilde{T}_{wa}$ is obtained by:
\begin{equation}
\tilde{T}_{wa} = T_{wh_j} T_{h_ia}, 
\end{equation}
where $T_{wh_j}$ and $T_{h_ia}$ are precomputed.

\subsection{Pose Clustering}

The vote set is usually big and very noisy (e.g., thousands of outlier votes for only a few dozen correct votes). Hence, votes are processed to find an accumulation somewhere, characterizing a convergence around a consensus value. Traditionally, in very low dimensions, the voting space is discretized into bins, each counting the number of votes received \cite{ballard_generalizing_1981}. Then the result is smoothed, and the maximum is found. However, the more the dimension of the space increases, the more the bins are numerous (and mostly empty), making the method unusable in practice. An alternative that scales better \cite{buch_rotational_2017} is to loop through votes and search for accumulation in their respective neighborhood. This idea is, for instance, the general idea used by DBSCAN clustering \cite{ester_density-based_1996}, which is well known to be fast and robust to outliers \cite{schubert_dbscan_2017}.

Using such an algorithm implies introducing a distance function on votes $(x, y, z, \psi) \in \mathbb{R}^3\times\left[0,2\pi\right]$ that defines what is a neighborhood of radius $\epsilon$. In particular, we must consider the different scale and periodic nature of the yaw $\psi$. This vote space can be seen as a subset of $SE(3)$ on which a frequent approach is to separate the contribution of the rotation from that of the translation and scale the rotation part by a factor $r$, often chosen experimentally \cite{bregier_defining_2018} :
\begin{equation}\label{eq:se3_distance}
d(T_1, T_2) = \sqrt{(\Delta\psi)^2 + r \lVert t_2 - t_1 \rVert^2},
\end{equation}
with $T_1$ and $T_2$ two gravity transform, $t_1$ and $t_2$ their translation part and $\Delta\psi$ the yaw angle difference in $[-\pi, \pi)$.

Besides, fully computing the distance matrix between each pair of (thousands of) votes to retrieve the $\epsilon$ neighborhoods would be intractable. Therefore efficient range search through spatial indexation (e.g., by a k-d tree) is needed. Such indexing is quickly built after the unfavorable periodic vote space is mapped into its 4-dimensional manifold image in $\mathbb{R}^5$:
\begin{equation}
f: \left(x,y,z,\psi \right) \mapsto \left( x,y,z,r.\text{cos}(\psi),r.\text{sin}(\psi) \right).
\end{equation}
For small neighborhoods of size $\epsilon \ll r$, the $SE(3)$ distance (\ref{eq:se3_distance}) defined on the vote manifold can be conveniently approximated by the Euclidean distance on $\mathbb{R}^5$.

\subsection{Pose Estimation}

After the clustering step, if no cluster is present, it implies that the Hough transform did not converge, which can mean that the input trajectory is outside the map. If there is at least one cluster, the best one is selected. For simplicity, it is defined as the cluster with the largest number of elements. Estimation is done by computing its centroid in $\mathbb{R}^5$, which is then projected back into vote space:
\begin{equation}
(x,y,z) = \frac{1}{|I|} \sum_{i \in I} (x_i, y_i, z_i), 
\end{equation}
\begin{equation}
\psi = \text{arctan}2\left(\frac{1}{|I|} \sum_{i \in I} \sin(\psi_i), \frac{1}{|I|} \sum_{i \in I} \cos(\psi_i)\right),
\end{equation}
with $I$ the index set of the selected votes, and $\lvert I \rvert$ its cardinality.

\section{EXPERIMENTAL EVALUATIONS}

The proposed magnetic Hough transform algorithm is evaluated regarding pose error, recall, precision, and runtime in 2D and 3D environments, including large open areas, staircases, and narrow corridors. A comparison with a state-of-the-art magnetic place recognition algorithm based on particle filtering \cite{solin_terrain_2016} is also presented.

\subsection{Datasets}\label{sec:dataset}

\begin{figure}
    \centering
    \includegraphics[trim=20 20 20 20,clip,width=0.96\linewidth]{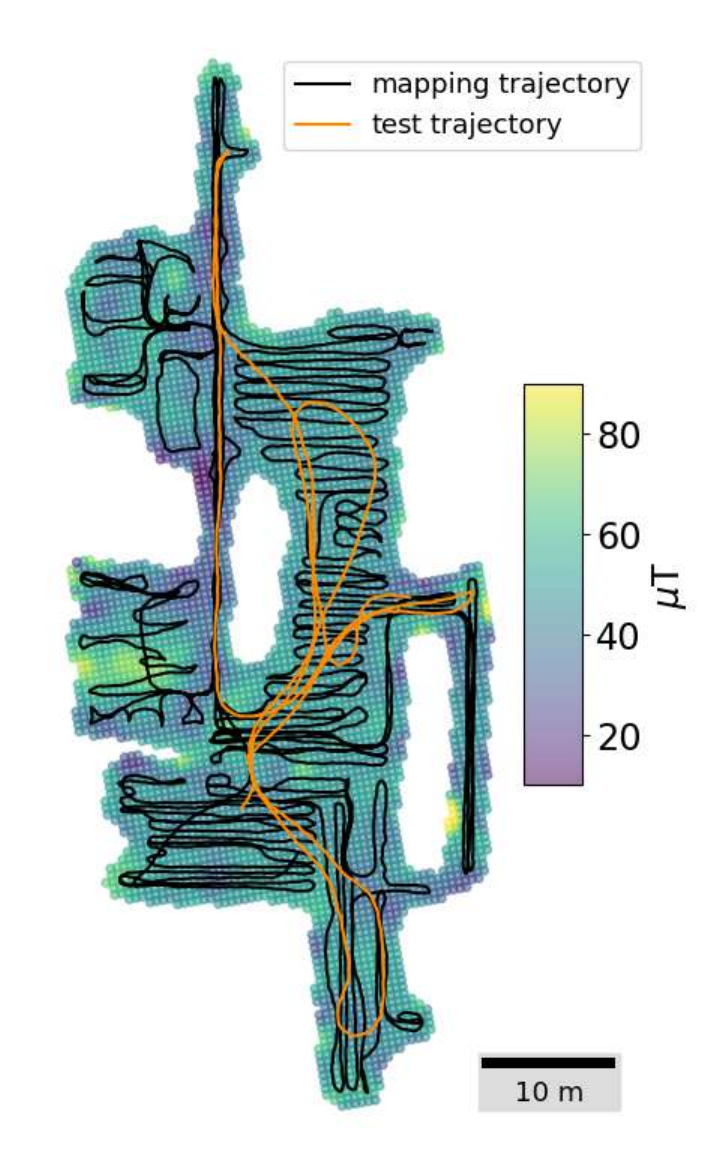}
    \caption{Top view of the \textit{atrium} dataset showing the mapping trajectory (black) and the test trajectory ground truth (orange). Note that the test trajectory does not follow the mapping path. Map is displayed using the magnetic field norm.}
    \label{fig:2Ddataset}
\end{figure}

\begin{figure}
    \centering
    \begin{subfigure}{.46\textwidth}
        \centering
        \includegraphics[trim=8 18 65 20,clip,width=0.98\linewidth]{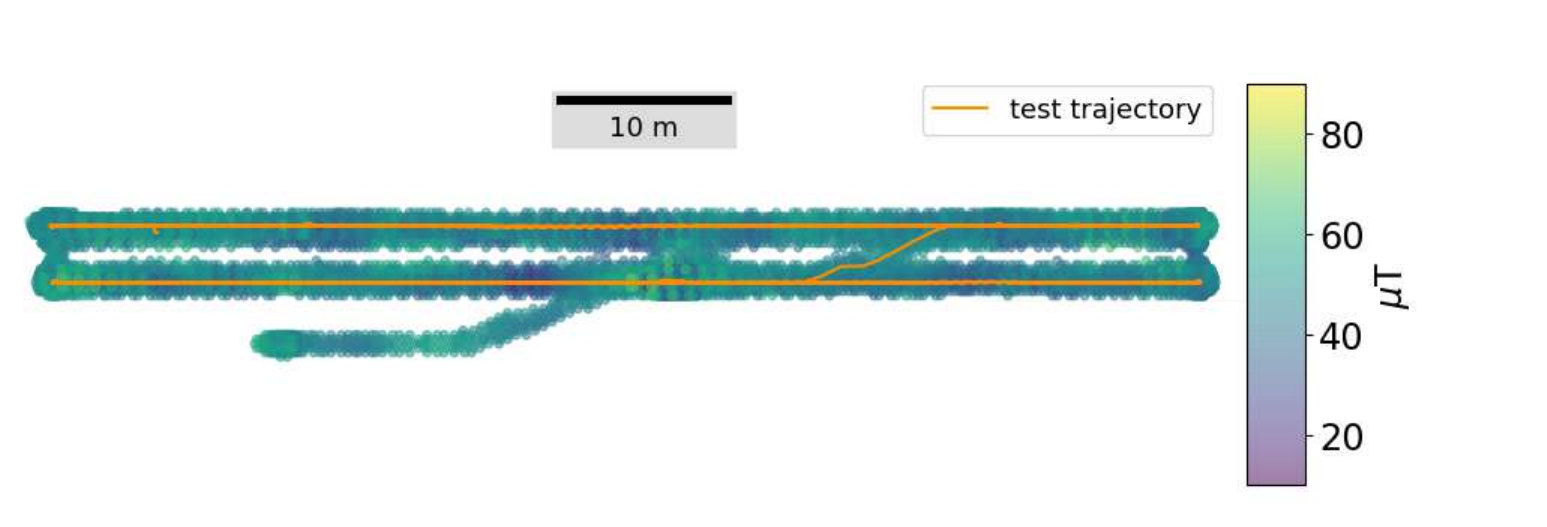}
        \caption{Side view}
        \label{fig:3Dtopdata}
    \end{subfigure}
    \begin{subfigure}{.46\textwidth}
        \centering
        \includegraphics[trim=30 30 30 90,clip,width=0.98\linewidth]{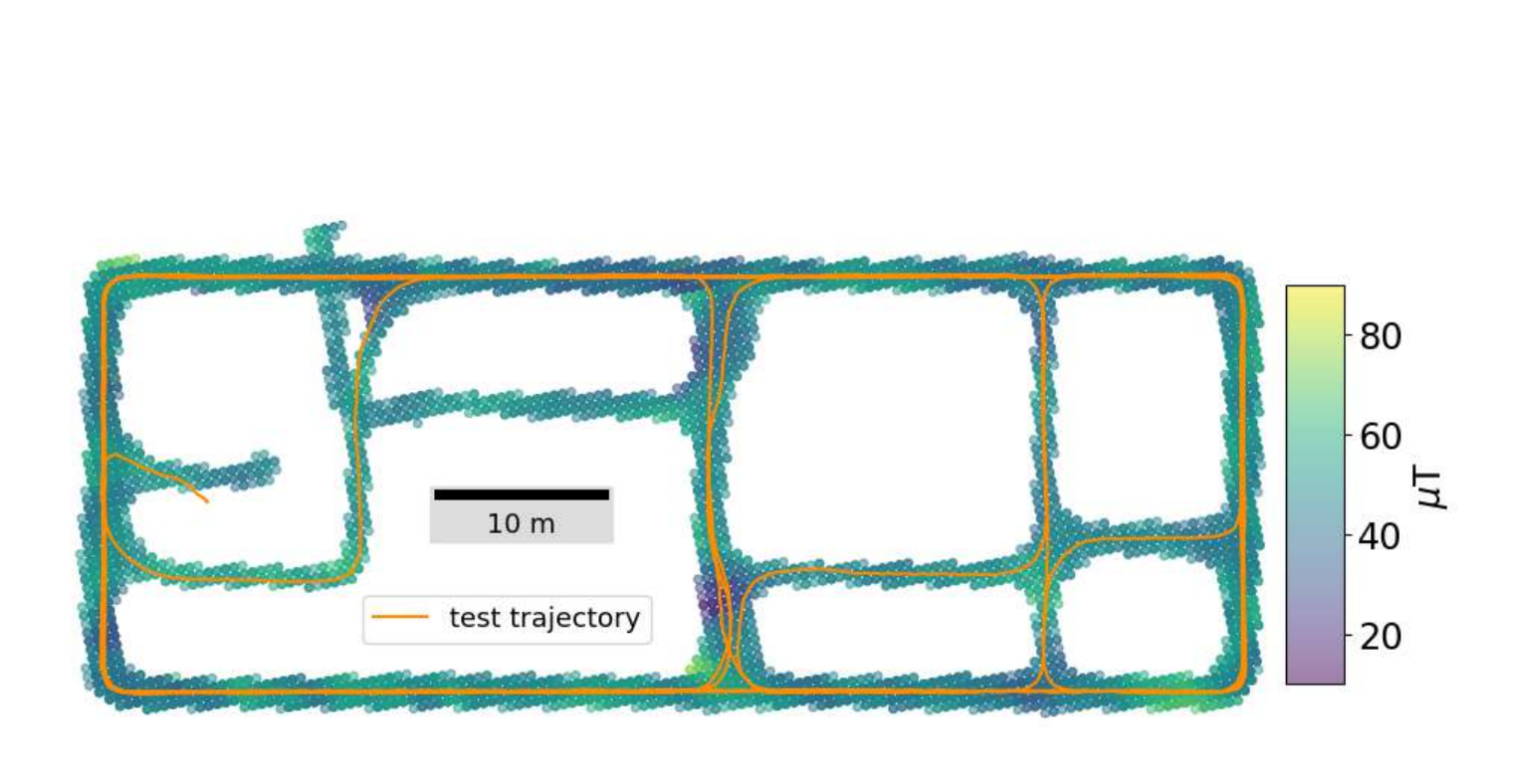}
        \caption{Top view}
        \label{fig:3Dsidedata}
    \end{subfigure}
    \caption{Visualization of \textit{corridor} dataset showing the 3d map of the building magnetic field magnitude and the test trajectory.}
    \label{fig:3Ddataset}
\end{figure}

\subsubsection{Hardware and sequences} To evaluate the performances of the MagHT algorithm in a challenging context, the test sequences (from which the input trajectories are extracted) have to not all follow the mapping path, and not all be acquired just after the mapping acquisition. Furthermore, different environments, such as open areas, staircases, and narrow corridors, must be covered. All sequences should also be associated with inertial data to compute the gravity constraint. To our knowledge, no public dataset combines all those characteristics. Hence we produced our own. Our acquisition platform is a helmet equipped with 4 FLIR Blackfly S cameras and an SBG-Ellipse-N containing an IMU and a magnetometer. All sensors are rigidly mounted and synchronized. Cameras and IMU are calibrated using Kalibr \cite{rehder_extending_2016}. The magnetometer is calibrated outdoors by spherical calibration \cite{renaudin_complete_2010}. However, it could be calibrated indoors too \cite{coulin_online_2022}. 

The first dataset called \textit{corridor} covers 1400 $\meter^3$ of corridors and staircases in a three-story building. The test sequence is one kilometer long and was acquired one year after the mapping sequence. The second dataset called \textit{atrium} covers 1200 $\meter^2$ of a ground floor. It includes a large atrium mapped with a trajectory deliberately different from the 240-meter test trajectory (see Fig \ref{fig:2Ddataset}). There, mapping and test trajectories often cross at large angles (even perpendicularly), and the goal is to assert that relocalizations rely only upon the map of the atrium, independently from the path taken to map it. Several hours separate both acquisitions.  

\subsubsection{Ground truth}\label{sec:gt} Ground truth poses are required for both building the magnetic map and quantitatively evaluating the accuracy of magHT. A visual-inertial graph SLAM \cite{campos_orb-slam3_2021} obtains them, with loop closure followed by a global bundle adjustment as a post-process. It also results in visual maps composed of 3d point clouds that can be used to express the ground truth poses of the test trajectories in the map coordinate frames. A visual relocalization algorithm is used to match 3D points between the two maps, and a 3D transformation is estimated from those 3D-3D correspondences by a Ransac algorithm. It typically yields a trajectory of poses located within a 10 centimeters error margin.

\subsubsection{Magnetic map} For a fair comparison, we used the same reduced rank Gaussian process extrapolation model \cite{solin_modeling_2018} for MagHT and the concurrent particle filtering algorithm \cite{solin_terrain_2016}. It is a continuous model that allows requests everywhere around the mapping sequence. However, our method is neither restricted to Gaussian processes nor continuous models and does not requires probabilistic values. Maps parameters defined in \cite{solin_modeling_2018} are fixed experimentally to sensible values describing typical variations: $\sigma_{\mathrm{lin}}^2, \sigma_{\mathrm{SE}}^2, \sigma_{\mathrm{noise}}^2 = 650,200,2  \,\mu\tesla^2$ and $l_{\mathrm{SE}}^2 = 1.69 \, \meter^2$ with 512 basis functions by domain of size $5\times5\times2 \,\meter^3$.

\subsubsection{Input trajectories} They are extracted from the test sequences. Using ground truth odometry (from the section \ref{sec:gt}) would be unrealistic for many applications. Thus the test sequences are also processed with monocular visual-inertial SLAM \cite{campos_orb-slam3_2021} without loop closure or any post-process. For the "corridor" dataset, 460 input trajectories of length 12 meters are extracted from a sliding window on the complete sequence. For the "atrium" dataset, roughly 120 input trajectories of each length 3, 6, 9, 12, 15, and 18 meters are extracted. All of them are first expressed in an arbitrary gravity frame before being processed by the proposed algorithm. In total, MagHT is tested on more than a thousand input trajectories.

\subsection{MagHT setup}\label{sec:maghtSetup}

\subsubsection{Map preprocessing} Offline, the continuous map described in section \ref{sec:dataset}  is discretized on a grid using steps of size $\lambda = 0.5 \,\meter$. However, in the 3D case, this grid is not cubic, as it would harm the clustering results. Indeed, because of the gravity assumption and because the input path is often at a constant walking altitude, a grid discretization in $z$ yields the same discretization in $z$ inside the voting space. This voting artifact can break one big cluster into multiple smaller ones for each $z$ layer. To reduce two fold the discretization step while keeping the number of element constant, half the column of a cubic grid are shifted upward by $\frac{\lambda}{2}$, resulting in a body-centered orthorhombic Bravais lattice of parameter $(\lambda\sqrt{2},\lambda\sqrt{2},\lambda)$.

\subsubsection{Input trajectories preprocessing} To not generate too many votes that would merge clusters, and because the magnetic field variations are low enough, measurements are smoothed to reduce noise and then downsampled using the same step length $\lambda$ as for the map sampling. Unlike a temporal one, spatial sampling is robust when the carrier is not moving. The origin of $a$ is translated at the barycenter of each input trajectory (not necessarily on the trajectory) to reduce voting errors induced by the lever-arm effect.

\subsubsection{Parameters} To illustrate MagHT's ease of use, we set the same parameter values on all datasets. Following DBSCAN parameter guidelines \cite{schubert_dbscan_2017}, \texttt{minpts} = 8, which is twice the vote DoF. Its range query size $\epsilon$ in the voting space should be as small as possible to reduce runtime, but it should be larger than the vertical $\frac{\lambda}{2}$ discretization. $\epsilon = \lambda$ gave good results. The distance function scaling factor $r$ is set to $5\,\meter$ such that the final translation error is approximately equal to the scaled yaw error. Matching in the feature space is done through the adaptative range:
$$
\delta = \min(\alpha\lvert m_i^a - m_{i-1}^a \rvert, \alpha\lvert m_{i+1}^a - m_i^a \rvert, \delta_{\max}) ,\quad \alpha \in \left[0,1\right], 
$$
with $i$ the index of the current input measure to match. Experimentally, $\delta_{\max} = 3.0 \,\mu\tesla$ and $\alpha = 0.67$. Notice that $\delta$ tends toward zero in uniform fields, which should avoid meaningless associations in outdoor environments.

\subsection{Particle filter setup}\label{sec:PfSetup}

We compared MagHT against a bootstrap particle filter \cite{solin_terrain_2016} as it is one of the only state-of-the-art methods to perform magnetic relocalization on trajectories that do not follow the mapping path. Hence its performances are used as a baseline. Because no open source code is available, we made our best effort C++ implementation, following as closely as possible their algorithm description, which is tailored for 2D navigation. In detail, the state is defined as $X = (x,y,\psi)$. At each timestep, the monocular visual-inertial odometry (VIO) projected in 2D replaces their pedestrian dead reckoning model in the dynamic equation. A stratified resampling is used jointly with an Effective Sample Size resampling criteria set to a standard value of 0.5. Particle initialization is performed using the first magnetic vector to align the yaw angle $\psi_0$. As in \cite{solin_terrain_2016}, their likelihood is based on the yaw invariant features described in section \ref{sec:vote}. Convergence is decided by setting a one-meter threshold on the particle x and y standard deviation. It is 2 to 3 times smaller than the equivalent criterion used in \cite{solin_terrain_2016}. The more accurate VIO allows it and avoids ambiguity from multimodal estimation.

\subsection{Metrics} 

Five criteria are used to evaluate both methods: The translation error $\lVert \hat{t}_{wa} - t_{wa}\rVert_2$, the rotation error $\lvert\hat{\psi}_{wa} - \psi_{wa}\rvert$, the precision, the recall, and the runtime. The precision is defined from the number of correct convergences (below one-meter error) over the total number of convergences, and the recall is defined from the number of correct convergences over all possible ones (i.e., the number of trajectories in the mapped area). The particle filter is indeed a \textit{filtering} algorithm, so the best performances are obtained at the last measurement. Hence, for a fair comparison, the particle filter metrics are computed using directly the transformation from the last sensor frame to the world frame instead of $T_{wa}$.

\subsection{Experiments}

Several experiments are conducted to assess several aspects of MagHT. All runtimes are achieved on one core of an 11th Gen Intel® Core™ i7-11800H with 16 GB of ram and no GPU. First, our method is compared against the particle filtering (PF) approach in paragraph \ref{sec:xpMaghtVsPf}. Particle filtering is about choosing the suitable trade-off between runtime and estimation quality. In our experiment, it is run with either 1600 or 6400 particles which will be abbreviated as pf1600 and pf6400. The second experiment described in paragraph \ref{sec:xp3DandStability} evaluate MagHT 3D relocalization capabilities and the magnetic field stability over time. Our proposed method performances in open and closed areas are compared in paragraph \ref{sec:xpopenVsClose}. The fourth experiment evaluates MagHT Robustness against false positives in unmapped areas in paragraph \ref{sec:xpunmap}. Finally, we assess the no drift hypothesis in paragraph \ref{sec:xpodom}.

\subsubsection{MagHT versus particle filtering}\label{sec:xpMaghtVsPf}

\begin{figure}
    \begin{subfigure}{.24\textwidth}
        \centering
        \includegraphics[trim=75 140 520 160,clip,width=0.98\linewidth]{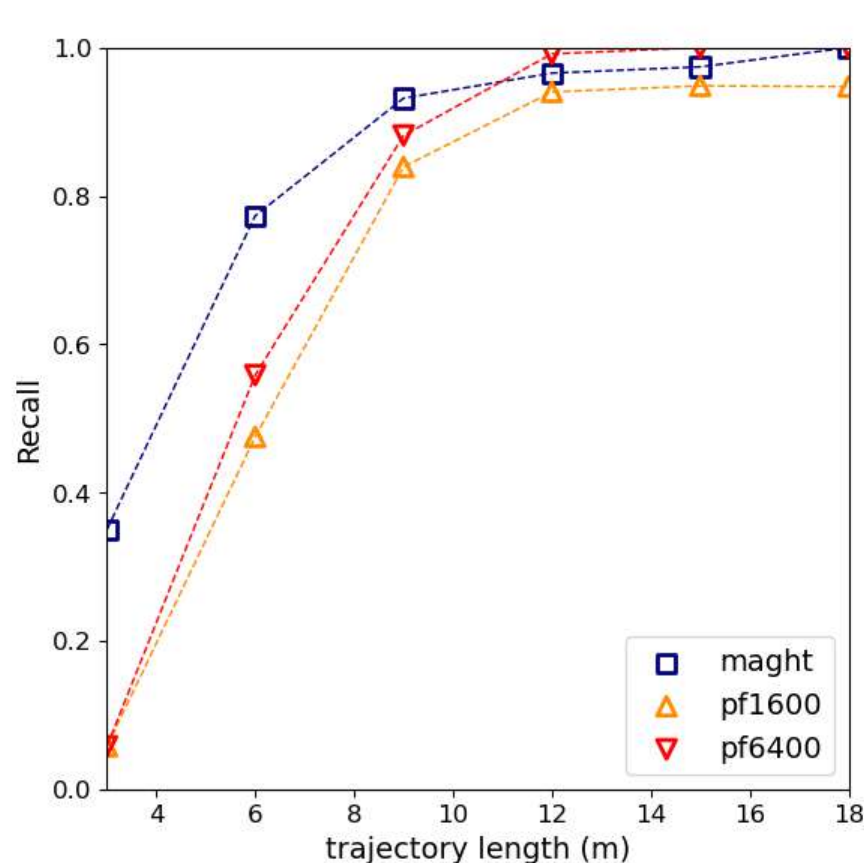}
        \caption{Recall}
        \label{fig:2Drecall}
    \end{subfigure}
    \begin{subfigure}{.24\textwidth}
        \centering
        \includegraphics[trim=510 140 85 160,clip,width=0.98\linewidth]{./images/pfvsmaght/Recall_and_Precision.pdf}
        \caption{Precision}
        \label{fig:2Dprecision}
    \end{subfigure}
    \begin{subfigure}{.24\textwidth}
        \centering
        \includegraphics[trim=75 140 520 160,clip,width=0.98\linewidth]{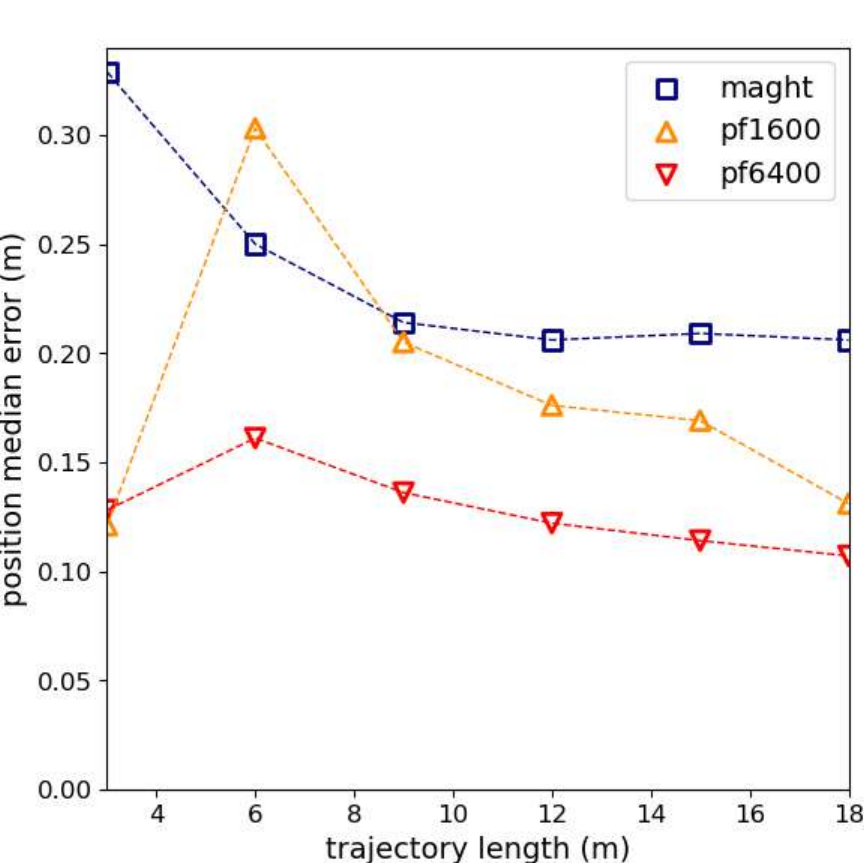}
        \caption{Median error (m)}
        \label{fig:2Derrpos}
    \end{subfigure}
    \begin{subfigure}{.24\textwidth}
        \centering
        \includegraphics[trim=75 140 520 160,clip,width=0.98\linewidth]{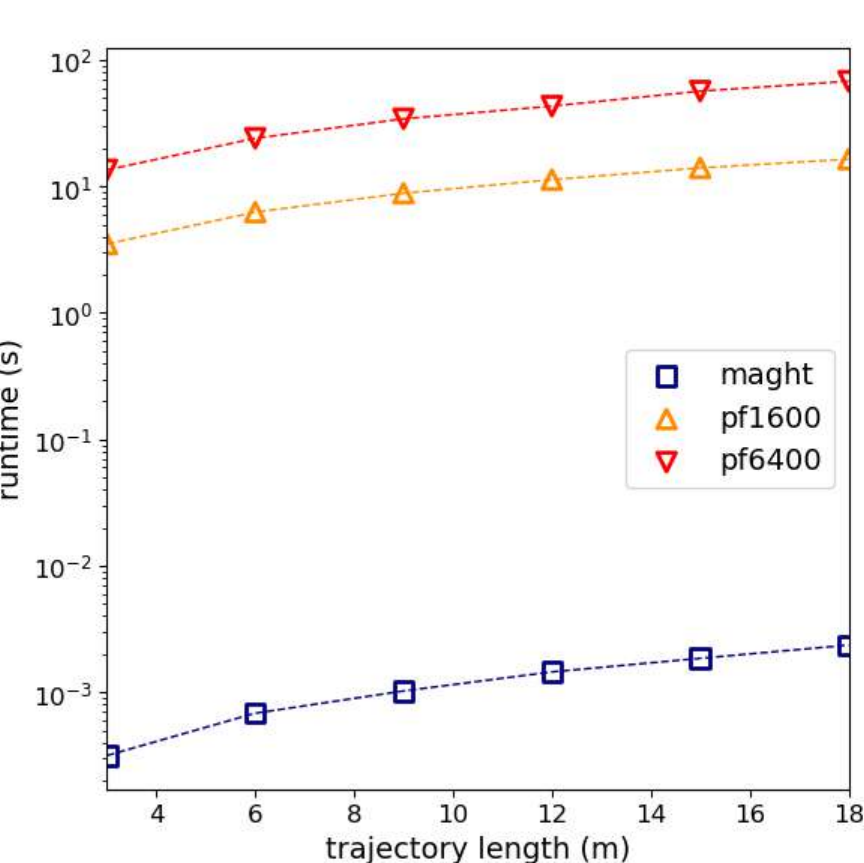}
        \caption{Runtime (s)}
        \label{fig:2Druntime}
    \end{subfigure}
    \caption{MagHT against particle filtering using 1600 and 6400 particles on variable trajectory lengths. In (d), odometry runtimes are not included.}
    \label{fig:pfvsmaghtstat}
\end{figure}

In this section, we only compare MagHT against a particle filter on the \textit{atrium} dataset, using input trajectory of variable lengths 3, 6, 9, 12, 15, and 18 meters. The 3D dataset \textit{corridor} is not used because the concurrent particle filter was initially proposed in 2D only. Besides, increasing the dimension state implies increasing the number of particles accordingly. One would expect similar results to the ones presented here but with worse computation time. Our method performances on \textit{corridor} are studied in paragraph \ref{sec:xp3DandStability}. For this experiment on \textit{atrium}, all metrics are displayed in Fig. \ref{fig:pfvsmaghtstat}. MagHT recall is superior on trajectories shorter than 9 meters. It stems from the fact that we had to increase the measurement noise of the filter to preserve the particle diversity and, therefore, reduce false positives. Indeed, \textit{filtering} algorithms that process measures one by one are more sensitive to outliers, especially at the beginning of the trajectory. However, higher noise values also increase the convergence time, thus reducing recall on shorter sequences. On the contrary, our algorithm processes all measures independently from their acquisition order, which increases its robustness. From this perspective, it is closer to the \textit{smoothing} algorithm family. Particle smoothing also exists \cite{sarkka_bayesian_2013} but suffers from even higher runtimes. For trajectories above 12 meters, MagHT recall is between pf1600 and pf6400. Fig. \ref{fig:2Dprecision} shows that the three algorithms have good precision, particularly pf6400, which has near zero false positives on all trajectory lengths. On the contrary, pf1600 precision is slightly lower than our proposed method. Sometimes no particles are near the correct solution, so it might converge toward a place where the magnetic field is similar. One can expect this effect to be magnified for larger environments. As visible in Fig. \ref{fig:2Derrpos}, MagHT errors are almost as good as the particle filtering ones. On 12 meter trajectory, median errors for our method, pf1600 and pf6400, are respectively 0.21, 0.18, and 0.12 meters. When the thousands of particles converged correctly, the sampling was dense around the true pose, which approximate almost perfectly its distribution. On the contrary, magHT usually has between ten and one hundred votes inside the most significant cluster. Nevertheless, it stills allows to average singular votes $\tilde{T}_{wa}$ by computing the cluster centroid $\hat{T}_{wa}$. The accuracy of $\hat{T}_{wa}$ might be improved by non-linear refinement post-processing since our method efficiency leaves enough computational power.

Overall, MagHT performances are similar to the particle filtering ones, but it is ten thousand times faster (see fig \ref{fig:2Druntime}). The main difference is that input and map data are associated with the feature space. Consequently, all magnetic field-related variables can be precomputed (e.g., the feature tree). Additionally, several online steps of our method are suitable for optimization through proper indexation trees. Conversely, the particle filter requires position and orientation hypothesis (particles) to compare a magnetic measure with the map. The larger the environment is, the more numerous the particles should be.
Furthermore, each comparison implies a map call. For 6400 particles on a 12-meter trajectory, it causes about 80 000 calls (16 times what MagHT needs). None of them can be precomputed as particles' positions are unpredictable beforehand. For all these reasons, MagHT runtimes outperform particle filtering ones by far.

\subsubsection{3D relocalization and map stability}\label{sec:xp3DandStability}

\begin{figure}
    \begin{subfigure}{.24\textwidth}
        \centering
        \includegraphics[trim=80 130 520 155,clip,width=0.98\linewidth]{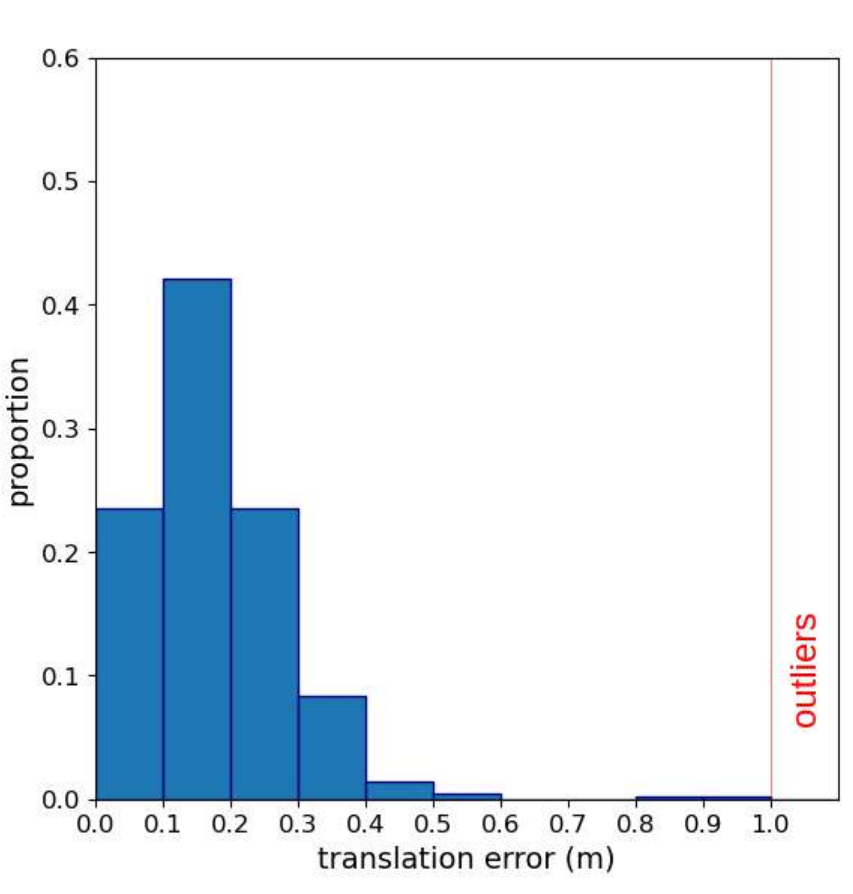}
        \caption{translation error on \textit{corridor}}
        \label{fig:3Derrorpos}
    \end{subfigure}
    \begin{subfigure}{.24\textwidth}
        \centering
        \includegraphics[trim=515 130 85 155,clip,width=0.98\linewidth]{./images/hist/histerrornofp3D_v3.pdf}
        \caption{yaw error on \textit{corridor}}
        \label{fig:3Derroryaw}
    \end{subfigure}
    \begin{subfigure}{.24\textwidth}
        \centering
        \includegraphics[trim=80 130 520 155,clip,width=0.98\linewidth]{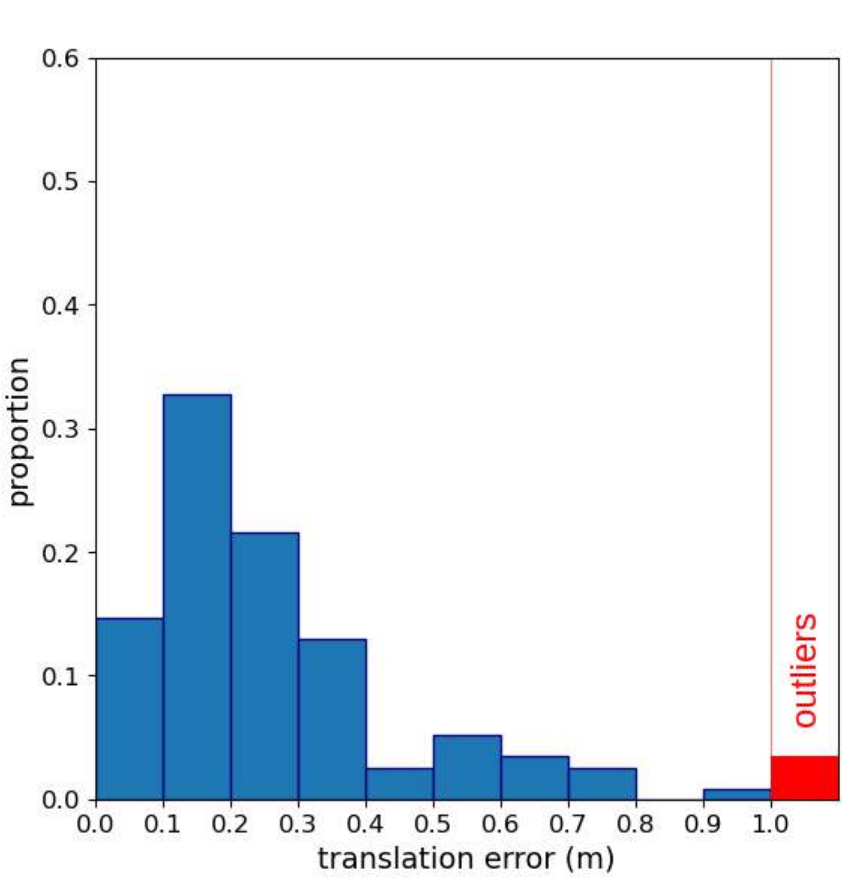}
        \caption{translation error on \textit{atrium}}
        \label{fig:2Derrorpos}
    \end{subfigure}
    \begin{subfigure}{.24\textwidth}
        \centering
        \includegraphics[trim=515 130 85 155,clip,width=0.98\linewidth]{./images/hist/histerrornofp2D_v4.pdf}
        \caption{yaw error on \textit{atrium}}
        \label{fig:2Derroryaw}
    \end{subfigure}
    \caption{Error histograms were computed using \textit{corridor} and \textit{atrium} on 460 and 120 input trajectories of length $12 \,\meter$, respectively. A threshold of 12$\degree$ defines false positives based on yaw errors. When scaled by the distance coefficient $r$, it is approximately equal to the one-meter threshold used for translation errors.}
    \label{fig:histogram_error}
\end{figure}

MagHT relocalization capability in a 3D environment as well as the magnetic field stability over time, are studied inside the \textit{corridor} dataset (see Fig \ref{fig:3Ddataset}). MagHT is executed on 460 input trajectories of length twelve meters \footnote{The previous experiment, in section \ref{sec:xpMaghtVsPf}, showed that twelve meters is the shortest length for best performances}. There is no false positive, and the recall is $89 \%$. Hence, it allows frequent and reliable relocalization. The error histograms are shown in Fig. \ref{fig:3Derrorpos} and \ref{fig:3Derroryaw}. Translation and yaw median errors equal $0.16 \,\meter$ and $1.08\,\degree$, respectively. These results confirm the magnetic field stability over time since the test sequence is acquired one year after the mapping acquisition. In addition to being accurate, the relocalization is lightweight, taking only $4.6\,\milli\second$. The offline 11000 features indexation in the k-d tree is also instantaneous (1.7 $\milli\second$), which is promising for an extension to online magnetic Simultaneous Localization And Mapping (SLAM). 

\subsubsection{Open area versus narrow corridors}\label{sec:xpopenVsClose}

In this section, we compare MagHT performances on the \textit{atrium} dataset with its performances on the \textit{corridor} dataset. Respectively 120 and 460 inputs trajectories of length 12 meters are extracted from both test sequences. Error histograms on those inputs are visible in Fig. \ref{fig:histogram_error}. Taking the \textit{corridor} dataset as a reference, recall increased from $89 \%$ to $97\%$, but the precision lowered from $100\%$ to $97\%$. Errors also increased from 0.16 to 0.21 meters and from 1.08 to 1.6 degrees. It stems from environment differences. First of all, in corridors, the distance to the closest ferromagnetic material is smaller than an atrium. Therefore, the field variations are higher in \textit{corridor}. Consequently, spatially close features might still be too different to be associated. It leads to a lower vote density and, therefore, to fewer clusters. Hence the recall is lower. On the contrary, the less variation the field has, the less observable the position is, which is not specific to our proposed solution. It partially explains the worse precision and errors in the atrium. Moreover, in open environments, the map is full. It further increases the vote density and, therefore, the risk of having large wrong clusters. The adaptive feature matching strategy still gave us good performances in both environments. Finally, notice that the high recall on \textit{atrium} demonstrates that MagHT is indeed able to perform relocalization even if the test trajectory crosses the mapping trajectory at high angles (see Fig. \ref{fig:2Ddataset}). Such angles would not be possible with sequence-based pattern-matching relocalization such as DTW.

\subsubsection{Robustness in unmapped area}\label{sec:xpunmap} 

This experiment assesses MagHT's robustness to relocalization attempts when the true pose is outside the map. Such a situation can appear when the environment is subject to change or when a robot may achieve exploration phases out of the previously mapped area. Hence we used the map of the \textit{corridor} dataset restricted to the second floor only, jointly with the input trajectories restricted to the third level. This way, over 250 input trajectories of 12 meters are in an unmapped area, and the map size is 400 $\meter^3$. MagHT yields only three false positives ($1.2 \%$). Each of the three false clusters contains exactly eight votes, the bare minimum given our DBSCAN parameters.

\subsubsection{Robustness to odometry inaccuracy}\label{sec:xpodom} 

Our algorithm does not model odometry errors which are unavoidable in real systems. Hence MagHT's sensibility to it should be characterized. To this end, the visual-inertial odometry on the \textit{atrium} dataset is computed using 1, 2, or 4 cameras without either loop closure or global bundle adjustment post-processing. After 12 meters, these three odometry systems yield median relative pose error (RPE) \cite{grupp2017evo} of (0.09 m, 0.3°), (0.05 m, 0.2°) and (0.04 m, 0.2°). Maght performances are not affected. The respective relocalization median errors are (0.21 m, 1.6°), (0.20 m, 1.6°), and (0.21 m, 1.5°) for the translation and yaw estimates. It confirms our hypothesis that the drift may be neglected on short input trajectories.

\section{CONCLUSION}

In this article, we introduced MagHT: a novel efficient, and accurate 3D relocalization algorithm based on the indoor magnetic field. To this date, other magnetic relocalization methods adapted to open areas are computationally expensive. On the contrary, our results demonstrate that MagHT execution time is several orders of magnitude faster than the concurrent particle filter while having comparable errors, recall, and precision. MagHT also performs fast relocalization in large 3D maps, which would require even more computational power for the filter-based method. 

MagHT opens several directions for research. In future work, we intend to integrate it in a graph SLAM framework, thus upgrading the current state-of-the-art, primarily based on DTW. Superior recall in open areas would allow for more loop closure and therefore reduced errors. Furthermore, MagHT's low false positive rate outside the mapped area would benefit most SLAM backends. Having no false positive at all is desirable and would require improving our method using additional rejection criteria, such as a ratio test on cluster sizes. The extension of MagHT to online magnetic mapping is possible since our experiment demonstrates that the precomputed feature tree is fast to build. Such trees are also editable efficiently. For this study, we collected the magnetic field by wearing our helmet system on our heads. Additional difficulties due to dynamic magnetic perturbations induced by a robot might be expected. We will better characterize its impact. Finally, building a coherent map from several maps can be challenging in a multi-agent SLAM context. Therefore, we will also extend MagHT to perform multi-magnetic map alignment. Generally speaking, MagHT efficiency makes it a good option for the localization of various embedded systems, evolving in large environments encompassed in non-uniform magnetic fields.

\addtolength{\textheight}{-12cm}   





\section*{ACKNOWLEDGMENT}

The author wishes to thank Fabrice MAYRAN DE CHAMISSO and Richard GUILLEMARD for their valuable feedback on early versions of the technique.

\bibliography{IEEEabrv,main}
\bibliographystyle{IEEEtran}

\end{document}